\documentstyle[12pt,epsf,rotate]{article}

\newcommand{\be}{\begin{equation}}
\newcommand{\ee}{\end{equation}}
\newcommand{\bea}{\begin{eqnarray*}}
\newcommand{\eea}{\end{eqnarray*}}
\newcommand{\ba}{\begin{eqnarray}}
\newcommand{\ea}{\end{eqnarray}}
\newcommand{\MSbar}{\overline{\rm MS}}
\newcommand{\lamFP}{\lambda_{\rm FP}}

\begin{document}

\begin{flushright}
DO--TH 96/20\\
TUM--HEP--253/96\\
October 1996
\end{flushright}
\vspace{1cm}

\begin{center}\large{\bf Matching conditions and\\ Higgs mass upper bounds
revisited.}\end{center}

\vspace*{15mm}

\begin{center}{Thomas Hambye$^*$
and Kurt Riesselmann$^{\dagger}$
\footnote{New address: DESY-IfH, Platanenallee 6, 15738 Zeuthen, Germany}
}\end{center}

\vspace*{5mm}

\begin{tabular}{l}
$^*$ Inst. f\"ur Physik T3, Universit\"at Dortmund, 44221 Dortmund, Germany
\\
$^{\dagger}$ Inst. f\"ur Theoretische Physik, Technische Universit\"at
M\"unchen, \\
\hspace*{5mm}85747 Garching b. M\"unchen, Germany
\end{tabular}

\vspace*{20mm}

\begin{center}{\bf Abstract} \end{center}
Matching conditions relate couplings to particle masses. 
We discuss the importance
of one-loop matching conditions in 
Higgs and top-quark sector 
as well as the choice of the matching scale. 
We argue for matching scales
$\mu_{0,t}\simeq m_t$ and $\mu_{0,H} \simeq$ max$\lbrace m_t, M_H\rbrace$.
Using these results, the two-loop Higgs mass upper bounds are reanalyzed.
Previous results for $\Lambda\approx$ few TeV are found to be too stringent.
For $\Lambda=10^{19}$ GeV we find $M_H < 180 \pm 4\pm 5$ GeV, the first error
indicating the theoretical uncertainty, the second error reflecting the
experimental uncertainty due to $m_t=175\pm6$ GeV.

\newpage

The standard model (SM) Higgs sector is usually considered to be an
effective theory. The possible triviality problem connected to the
underlying $\phi^4$ theory \cite{triviality} can be avoided if new
physics appears at some high energy $\Lambda$.  Depending on the
specific value of $\Lambda$, an upper bound on the mass $M_H$ of the
SM Higgs boson can be derived \cite{oldwork,lindner,grad}.  This upper
bound is connected to an unsatisfactory high energy behaviour of the
Higgs quartic self-coupling $\lambda$ if $M_H$ is large. It manifests
itself in the (one-loop) Landau pole \cite{landau} when using a
perturbative approach, or in large cutoff effects when performing
lattice calculations \cite{lattice,LW,aachen,heller}.

Previous work \cite{lindner,grad} extensively investigated the
dependence of the $M_H$ upper bound on the top quark mass $m_t$. The
discovery of the top quark and the steadily improving mass
determination of $m_t$ leads to the question: Which uncertainties are
remaining in the theoretical prediction of the $M_H$ upper bound?
Using the perturbative approach up to two loops, we investigate the
sensitivity of the $M_H$ upper bound with regard to various cutoff
criteria, the inclusion of matching corrections, the 
choice of the matching scale $\mu_0$, and the remaining top mass dependence. 

In Sect.~I we review the scale dependence of the SM matching conditions.
We argue that the most reasonable choice of the matching scale
for the Higgs quartic coupling is $\mu_{0,H} \simeq$ max$\lbrace m_t, M_H
\rbrace$, and the top quark Yukawa coupling should be fixed at $\mu_{0,t}
\simeq m_t$.  In particular, the use of the scale $M_Z$ leads to unreliable
results in the case of the Higgs coupling.  Using these observations, 
we recalculate the SM Higgs mass upper bounds at the two-loop level (Sect.~II).
At low cutoff scale, $\Lambda \simeq 10^3-10^4$ GeV, we find the theoretical
uncertainties to be large, $O(200$ GeV), even when using the choice 
$\mu_{0,H}=M_H$.  
At high cutoff scale, $\Lambda \simeq 10^{15}-10^{19}$ GeV, the
theoretical uncertainties are greatly reduced and amount to $O(10$ GeV). The
additional experimental uncertainty entering through $m_t$ can be neglected
except for very large cutoff scales.  Assuming a top quark mass of
$m_t=175\pm6$ GeV \cite{warsawMT} and a cutoff scale $\Lambda=10^{19}$ GeV, 
we find an upper bound $M_H < 180 \pm 4 \pm 5$ GeV, where the first error
estimates the theoretical uncertainty, and the second error indicates the top
quark mass dependence.

\newpage

\vspace{10mm}
\begin{center}{\bf I. Matching Conditions}\end{center}

We start with a detailed look at the so-called matching conditions in
the SM: the
relations between the physical masses and the corresponding 
running couplings.  This part of our letter is therefore not specific to the
calculation of Higgs mass upper bounds but has further applications.

The $\MSbar$ Higgs quartic coupling, $\bar{\lambda}$, and 
top Yukawa coupling, $\bar{g}_t$, are related to
$M_H$ and $m_t$ using the following matching conditions: 
\begin{eqnarray}
\label{mchiggs}
\bar \lambda (\mu_0) &=& \frac{M^2_H}{2v^2}  [1+\delta_H(\mu_0)]\,,
\\
\label{mctop}
{\bar g_t (\mu_0)} &=& \frac{\sqrt{2} m_t}{v} [1+\delta_t(\mu_0)]\,,
\end{eqnarray}
where $v = (\sqrt{2} G_F)^{-1/2} \approx 246$ GeV. 
The definitions of the 
corresponding tree level couplings are
obtained by dropping the matching corrections $\delta$,  thus
fixing our notation.
The use of two-loop renormalization group (RG) equations in connection
with $\MSbar$ couplings requires one-loop expressions for the
corrections $\delta_H(\mu_0)$ and $\delta_t(\mu_0)$; they are given in
\cite{sirlin} and \cite{HK}, respectively.

In the case of the electroweak gauge couplings, one-loop matching corrections
have also been calculated \cite{sir,sirmar}.  However, it is custom to extract
the $\MSbar$ gauge couplings directly using $\MSbar$ definitions for
experimental 
observables. The measured values for the $\MSbar$ electroweak mixing angle 
and QED coupling 
fix the $\MSbar$ electroweak couplings at the scale $\mu_0=M_Z=91.187$ GeV
\cite{PDG}:
\begin{eqnarray}
\label{alphamsbar}
{{\bar\alpha}^{-1}}(M_Z) &=& 4 \pi \frac{\bar{g}^2(M_Z)+\bar{g}'^{2}(M_Z)}
{\bar{g}^{2}(M_Z) \bar{g}'^{2}(M_Z)} =
127.90 \,,\\
\label{sinmsbar}
\sin^2\theta^{\MSbar}_W (M_Z)&=& 
\frac{\bar{g}'^{2}(M_Z)}
{\bar{g}^2(M_Z)+\bar{g}'^{2}(M_Z)} = 
0.2315 \,.
\end{eqnarray}
The $\MSbar$ electroweak couplings are obtained as $\bar
g(M_Z)=0.651...$ and $\bar g'(M_Z)=0.357...$.

For comparison, it is nevertheless interesting to define gauge sector
matching conditions 
in analogy to Eqs.~(\ref{mchiggs}) and
(\ref{mctop}), that is, using gauge boson masses and matching corrections:
$\bar{g}^2(M_Z) \equiv \frac{4 M_W^2}{v^2}(1+\delta_W)$ and 
${\bar{g}^2}(M_Z)+{\bar{g}'^2}(M_Z) \equiv \frac{4
M_Z^2}{v^2}(1+\delta_Z)$. Taking 
$M_W=80.35$ GeV
and $M_Z$ as above one obtains $\delta_W\approx-0.4\%$ and 
$\delta_Z\approx0.7\%$.
As we will see below, the one-loop matching corrections $\delta_t$ and
especially $\delta_H$ are significantly larger.

In the following we examine in detail the interesting structure of
the matching corrections $\delta_H(\mu_0)$ and $\delta_t(\mu_0)$ as a function
of $\mu_0$ and $M_H$. For $\delta_H$, the heavy
top mass of $m_t=175$ GeV changes drastically the original 
discussion\footnote{The analysis of \cite{sirlin} is based on a
value $m_t$=40 GeV.} 
presented in \cite{sirlin}, except for $M_H\gg m_t$.

\begin{figure}[h]
\vspace*{30pt}
\centerline{
\hspace{-8.0cm}\epsfysize=3.6in {\epsffile{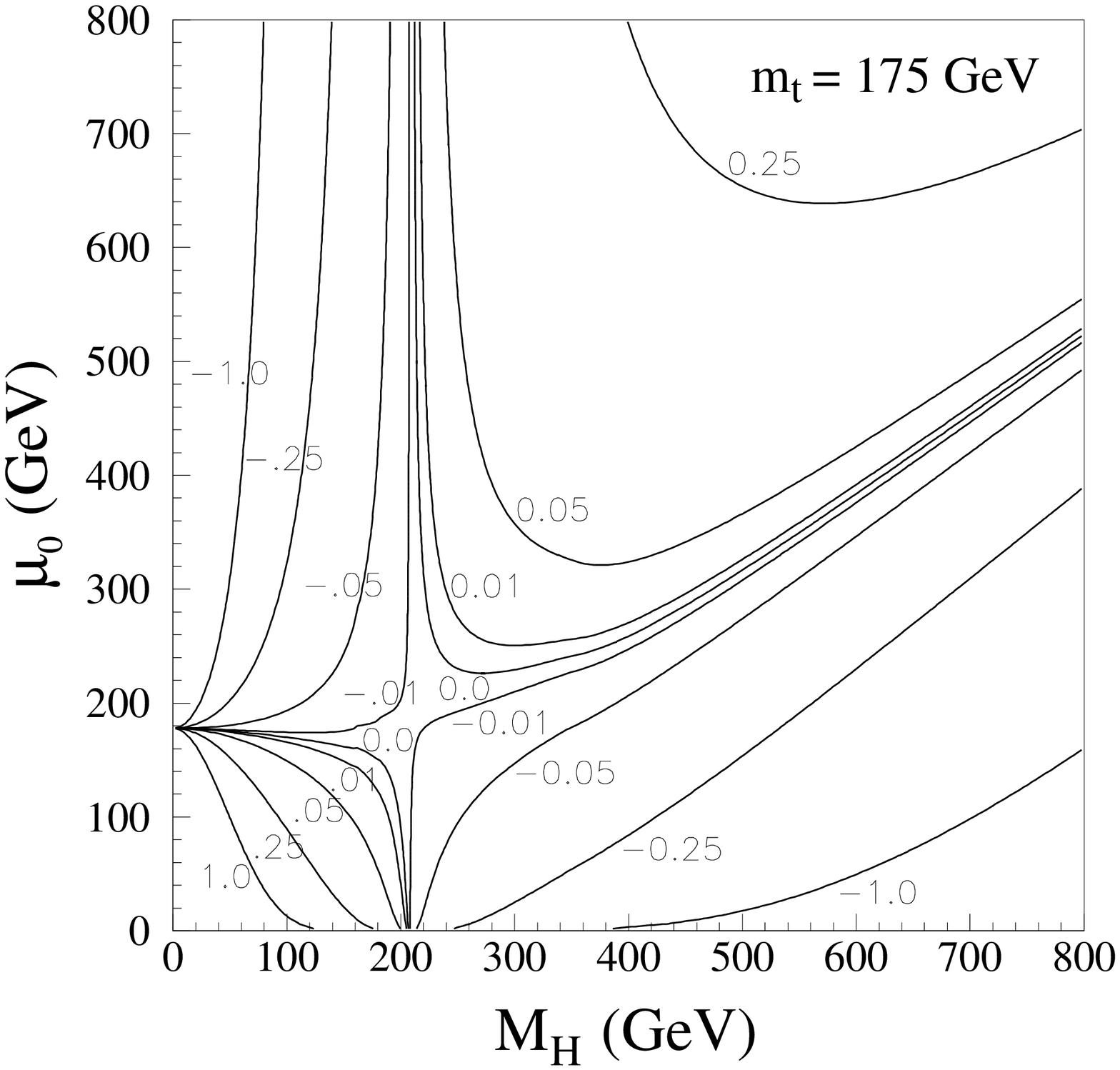}}
}
\vspace{0.5in}
\caption{(a) Values of $\mu_0$ and $M_H$ for which the one-loop Higgs matching
  correction $\delta_H(\mu_0)$, Eq.~(\protect\ref{mchiggs}), equals the
  values indicated next to the various contour lines. The top quark
  mass is taken to be 175 GeV. (b) Same plot, but the leading two-loop
  heavy-Higgs corrections \protect\cite{NR} have been added.}
\label{figdeltah}
\end{figure}

Using the result of \cite{sirlin}, the correction $\delta_H(\mu_0)$ 
can be rewritten in the following way:
\be
\delta_H(\mu_0)=\frac{2v^2}{M^2_H} \frac{1}{32\pi^2v^4}\{
h_{0}(\mu_0)+M^2_H h_1(\mu_0)+{M^4_H}h_{2}(\mu_0)\}
\label{deltahiggs}
\ee
with
\ba
\label{eqh0}
h_{0}(\mu_0) &=&
-24m^4_t\ln\frac{\mu^2_0}{m^2_t}+6M^4_Z\ln\frac{\mu^2_0}{M^2_Z}+12M^4_W \ln
\frac{\mu^2_0}{M^2_W} +c_{0}\,,\\
\label{eqh1}
h_1(\mu_0) &=&
12m^2_t\ln\frac{\mu^2_0}{m^2_t}-6M^2_Z\ln\frac{\mu^2_0}{M^2_Z}-12M^2_W \ln
\frac{\mu^2_0}{M^2_W} + c_{1}\,,\\
\label{eqh2}
h_2(\mu_0) &=& \frac{9}{2} \ln \frac{\mu^2_0}{M^2_H}+\frac{1}{2}
\ln\frac{\mu^2_0}{M^2_Z}+ \ln \frac{\mu^2_0}{M^2_W} +c_{2}\,.  \ea The
constants $c_i$ are independent of $\mu_0$. For $m_t$=175 GeV and 75
GeV $ < M_H <$ 570 GeV their total contribution to $\delta_H$ is in
magnitude less than 0.02, though some individual terms can exceed 0.05. 
Depending on the choice of $\mu_{0}$, the logarithmic terms in
Eqs.~(\ref{eqh0}) -- (\ref{eqh2}) can yield a much larger
correction. In Fig.~\ref{figdeltah}a we show the one-loop result of
$\delta_H$ as a function
of $\mu_{0}$ and $M_H$ for $m_t$ = 175 GeV. We find that 
the matching correction $\delta_H$
can be in magnitude larger than 25\% for various regions in the
parameter space $(\mu_{0},M_H)$, even exceeding 100\%.  Clearly the
matching correction should be taken into account and the
choice of the matching scale $\mu_{0}$ is important:
some choices are more appropriate than others.

To discuss the dependence of $\delta_H$ on $\mu_0$ we consider
its derivative:
\ba
\label{beta1lp}
\frac{d\delta_H(\mu)}{d\mu}\!\! &\!\!=\!\!&\!\!
\frac{1}{\mu} \frac{2v^2}{M^2_H}\frac{1}{8\pi^2v^4}
\left[3M^4_H
-3 M^2_H(M^2_Z+2M^2_W-2m^2_t)+3M^4_Z+6M^4_W-12m^4_t
\right] \nonumber\\
\!\!&\!\!\equiv\!\!&\!\! \frac{1}{\mu} \frac{2v^2}{M^2_H} \beta_\lambda \,,
\ea
where $\beta_\lambda$ is the one-loop beta function of the coupling $\lambda$
expressed in terms of the different physical masses
rather than in terms of the various $\MSbar$ 
couplings (which is consistent at
one-loop order). For $m_t=175$ GeV, $\beta_\lambda$ 
equals zero if $M_H\simeq208$ GeV.
Taking $M_H$ to be different from this value, $\beta_{\lambda}$ quickly
becomes large.  
If $M_H \ll$ 208 GeV,  the $m^4_t$ contribution 
dominates and $\beta_\lambda \ll 0$; 
if $M_H \gg$ 208 GeV,  the $M^4_H$ contribution
causes $\beta_\lambda \gg 0$.
Correspondingly, the magnitude of $\delta_H$ is insensitive 
to the choice of
$\mu_{0}$ only for a small range of $M_H$ values, see Fig.~\ref{figdeltah}.

Natural choices of $\mu_0$ in Eq.~(\ref{mchiggs}) are the various
masses appearing in the logarithms in Eqs.~(\ref{eqh0})--(\ref{eqh2}):
$M_H$, $m_t$ or $M_Z$.  Since the impact of the choice of $\mu_{0}$
is connected to the value of $M_H$, we consider three cases:\\
(1) \underline{$M_H\ll m_t$ ($M_H\simeq$ 70-100 GeV)}: This is the
range where $\beta_\lambda\ll 0$ due to the dominant $m_t^4$ term in
Eq.~(\ref{beta1lp}). Such a large contribution to $\beta_{\lambda}$ is
possible for low values of $M_H$ because there is no symmetry in the
scalar sector which imposes $\beta_{\lambda}$ to go to 0 for
$\lambda\rightarrow 0$. This is in contrast to the beta functions of
the gauge and Yukawa sectors.  Consequently, 
the coefficients of the logarithmic terms in
$\delta_H$ can be large for
small values of $M_H$, actually going to infinity as $M_H\rightarrow 0$.
(In contrast, the coefficients in the matching corrections of the non-scalar
sectors vanish or approach a finite constant if the 
corresponding particle mass goes to
zero). 
Indeed, the $m^4_t$ term in $\beta_{\lambda}$ gives rise to
the large coefficient $m^4_t/M^2_H$ which multiplies
$\ln(m^2_t/\mu^2_0)$. (The overall factor $1/M^2_H$
is present because $\delta_H$ is the ratio of the loop contribution to
the lowest order contribution to $\bar{\lambda}$, the latter being
proportional to $M^2_H$).  Consequently if $M_H$ is small then
$\delta_H$ is small only if $\mu_0$ is chosen close to $m_t$, not $M_H$. 
For example, if
$M_H = 70$ GeV and $m_t=175$ GeV, we find $\delta_H(M_H)\simeq80$\%
whereas $\delta_H(m_t)\simeq-0.7$\%.
The dominance of the $\ln(\mu_0^{2}/m_t^2)$ term 
indicates that the top mass scale is the correct scale of
reference for low values of $M_H$. 
Interestingly, even if the top
one-loop correction to $\delta_H$ is large, perturbation theory is
still applicable : $\delta_H$ is formally the product of a series in
powers of $g_t$ and $\lambda$, with an overall factor 1/$M^2_H$.  The
higher-order terms contributing to $\delta_H$ are expected to be small 
in the same way in which the
two-loop term of $\beta_\lambda$ \cite{einhorn,vaughn2} is smaller
than the one-loop contribution to $\beta_\lambda$.\footnote{
The simultaneous largeness and perturbativity of the top
quark contribution in the scalar sector could be the origin of the
symmetry breaking of SU(2) $\times$ U(1). A recent model \cite{fatelo}
using this approach yields $M_H\simeq$ 80-100 GeV.}
\\
(2) \underline{$M_H \simeq$ 0.8-1.7 $m_t$}: Taking $m_t=175$ GeV, the
function $\beta_{\lambda}$ features a zero in this Higgs mass
range, indicating that both Higgs and top-quark contributions have similar
weight. Both $\mu_0=m_t$ and $\mu_0=M_H$ are acceptable choices.
In fact, we find the Higgs matching corrections to satisfy 
$|\delta_H| < 5\%$ for a large range of $\mu_0$ around
$\mu_0 \simeq$ $M_H \simeq m_t$.  
This property remains true if
the top quark mass has a value somewhat different from $m_t=175$ GeV.
Choosing $\mu_{0}=$ max$\lbrace m_t, M_H\rbrace$,
a variation of 160 GeV $<m_t<$ 190 GeV results
in $-1.1\%<\delta_H(m_t)< -1.0\%$ if $M_H=140$ GeV, and
2.4$\%$ $<\delta_H(M_H)<$ 3.6$\%$ if $M_H=300$ GeV.\\ 
(3) \underline{$M_H \gg m_t$.}: Such a value of $M_H$ causes a large
and positive value of $\beta_{\lambda}$. The leading logarithmic
contribution to $\delta_H$ is the $M^2_H\ln(\mu^2_0/M^2_H)$ term which
can be suppressed choosing $\mu_0 \simeq M_H$. Yet the other
terms, including $\ln(\mu^2_0/m^2_t)$, are viable 
for $\mu_0=M_H$. For example, $M_H= 570$ GeV results in
$\delta_H(M_H)\simeq 20\%$.
For larger $M_H$ the matching correction approaches the heavy-Higgs result
\begin{equation}
\delta_H=\frac{M_H^2}{32\pi^2v^2}\left( 12\ln\frac{\mu_0^2}{M_H^2} 
+ 25-3\pi\sqrt{3}\right)\,.
\end{equation}
A possible choice, used in \cite{sirlin}, would be  $\mu_0 \simeq 0.7 M_H$ 
such that $\delta_H(\mu_0)\simeq 0$.
This approach, however, fails at two
loops since the two-loop heavy-Higgs terms are sizeable \cite{NR}.
Adding these two-loop contributions to the full one-loop result of $\delta_H$,
we show the resulting $\mu_0$ dependence in Fig.~\ref{figdeltah}b.
A satisfactory perturbative behaviour is obtained for $\mu_0=M_H$ if 
$M_H<O(800$ GeV). The choices $\mu_0=m_t$ or $M_Z$ are inappropriate since
they lead to unreliable
perturbative predictions for even smaller values of $M_H$.  In particular,
the choice $\mu_0=M_Z$ leads to $\delta_H< -1.0$ for $M_H>690$ GeV, resulting
in an unphysical {\it negative} $\MSbar$ Higgs coupling.
\\ 
Summarizing our results for the three different Higgs-mass scenarios described
above, we find 
the scale $\mu_{0} \simeq$ max$\lbrace m_t, M_H \rbrace$ to be the appropriate
Higgs matching scale for $m_t \simeq 175$ GeV.
The calculation of the $M_H$ upper bound (see Sect.~II) is an example
how physical quantities are sensitive to the choice of $\mu_0$.

Next we consider the matching correction 
$\delta_t(\mu_0)$ entering Eq.~(\ref{mctop}).
It has been given 
at one loop in \cite{HK}, with the dominant QCD correction
given earlier in \cite{deltatQCD} and the Yukawa corrections in 
\cite{deltatYuk}. The result can be written as
\be
\delta_t(\mu_0)=\left(
-4\frac{\alpha_s}{4\pi}
-\frac{4}{3}\frac{\alpha}{4\pi}
+\frac{9}{4}\frac{m^2_t}{16\pi^2v^2}\right)\,\ln\frac{\mu^2_0}{m_t^2}
\;+\;c_t\,,
\label{deltatop}
\ee
where $c_t$ is independent of $\mu_{0}$ and can be evaluated 
using the results in \cite{HK}.\footnote{There is a misprint in 
Table I of \cite{HK}: the term
$6.90\times 10^{-3}$ should have the opposite sign.} 
Taking $\alpha_s(M_Z) = 0.118$ \cite{alphaqcd}, we 
find $-0.052 < c_t <
-0.042$ 
for top quark mass of 150 GeV $< m_t <$ 200 GeV and Higgs
mass of 50 GeV $< M_H <$ 600 GeV. 
The correction due to $c_t$ is
therefore in magnitude larger than the sum of the
$\mu_{0}$-independent contributions $c_i$ to $\delta_H$.  The
largeness of $c_t$ is mostly due to the QCD correction.  In contrast,
there is no one-loop QCD correction contributing to $\delta_H$.

\begin{figure}[tb]
\vspace*{30pt}
\centerline{
\epsfysize=5.0in {\epsffile{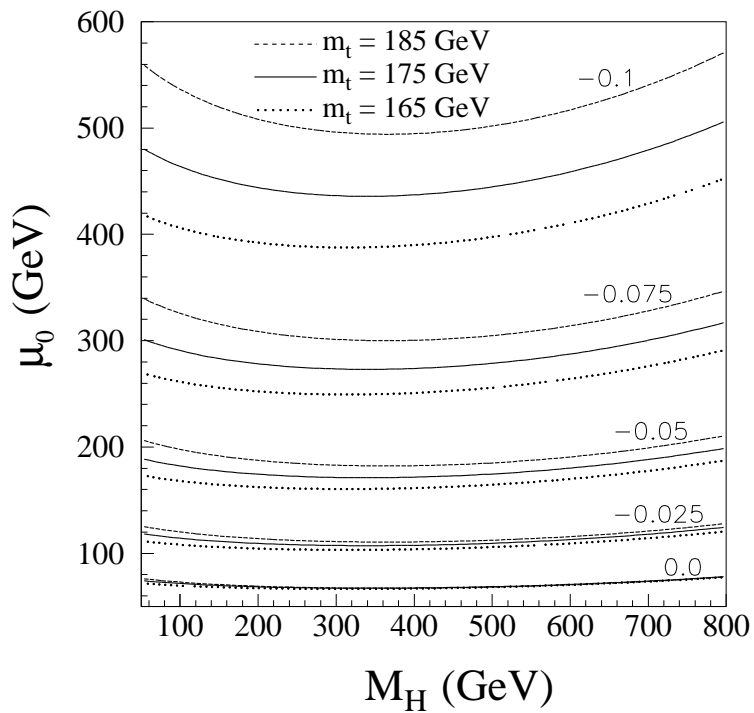}}
}
\vspace{0.5in}
\caption{Values of $\mu_0$ and $M_H$ for which top-quark matching
 correction $\delta_t(\mu_0)$, Eq.~(\protect\ref{mctop}), equals the values
 indicated next to the various contour lines. Results are shown using
 $m_t=165$ GeV (dotted), 175 GeV (solid), and 185 GeV (dashed).}
\label{figdeltatop}
\end{figure}

Since LEP I provides the result for $\alpha_s$ at scale $M_Z$, it seems
plausible to use a matching scale $\mu_{0} = M_Z$. 
This yields $\delta_t(M_Z) \simeq O(-2\%)$ as can be seen in
Fig.~\ref{figdeltatop}.  Looking at the logarithm appearing in
Eq.~(\ref{deltatop}), however, the adequate choice is $\mu_0 \simeq
m_t$: no other particle mass enters the $\mu_0$-dependent logarithms.
With this choice we immediately obtain $\delta_t(m_t) = c_t =
O(-5\%)$. Here the difference in taking $\alpha_s(M_Z)$ {\it vs.} 
$\alpha_s(m_t)$ amounts to higher-order corrections which are suppressed.
 
The present-day experimental result of $m_t=175\pm6$ GeV
\cite{warsawMT} leads to $\pm 3.4$\% uncertainty in the tree-level result of
$g_t$.  Comparing with the results above, 
we find the one-loop matching correction $\delta_t$ to be of equal importance.
This concludes our review of the matching conditions.

\vspace{10mm}
\begin{center} {\bf II. Higgs Mass Upper Bounds}\end{center}

The triviality problem of the SM is completely fixed by the beta
functions of the theory.  The functions $\beta_i$ for all SM couplings
have been calculated in the $\MSbar$ scheme up to two loops
\cite{einhorn,vaughn2,vaughn1}. At the 
one-loop level, a heavy Higgs particle gives rise to a positive
function $\beta_\lambda$, causing the running Higgs quartic coupling
$\lambda(\mu)$ to permanently increase as $\mu$ increases. At some
value $\mu=\Lambda_L$, the position of the one-loop Landau 
pole \cite{landau}, the
Higgs running coupling becomes infinite: perturbation theory has
ceased to be meaningful long before.

At the two-loop level, a heavy Higgs mass causes $\lambda(\mu)$ to
approach an ultraviolet (metastable) fixpoint. This fixpoint is almost
entirely determined by the leading Higgs coupling contributions to
$\beta_\lambda$ at two loops:
\begin{equation}
\beta_\lambda =    24\frac{\lambda^2}{(16\pi^2)^2}
                - 312\frac{\lambda^3}{(16\pi^2)^3}\,.
\label{betafct}
\end{equation}
The resulting fixpoint value, corresponding to $\beta_\lambda=0$, is
\begin{equation}
\lamFP = 12.1... \:\,.
\label{fixpoint}
\end{equation}
Increasing the scale $\mu$ even further, the growing value of the
running top quark coupling can no longer be neglected and changes
$\beta_\lambda$, hence modifying the above fixpoint behaviour. Since
perturbation theory is already meaningless even before $\lambda(\mu)$
reaches $\lamFP$, we are not concerned about the details of
the $\lambda(\mu)$ behaviour beyond the metastable fixpoint.

At three loops,
only the
leading contribution to $\beta_\lambda$ is known \cite{NR,LWbeta3lp,rus}. It
causes the running Higgs coupling to again have a Landau singularity.
Since the complete set of three-loop
SM beta functions and the corresponding two-loop matching conditions
are not yet available, we restrict our present analysis to two-loop beta
functions.

To obtain $M_H$ upper bounds from the RG evolution of $\lambda(\mu)$
to some embedding scale $\mu=\Lambda$, one has to choose a cutoff
value for $\lambda(\Lambda)$. We denote this cutoff condition by
$\lambda_c({\Lambda})$.
At one loop, the standard choice is to require
that $\lambda(\mu)$ avoids the Landau singularity for $\mu<\Lambda$.
This corresponds to $\lambda_c(\Lambda)=\infty$.  At two loops, the running
Higgs coupling remains finite and
$\lambda(\mu)\rightarrow\lamFP$ as $\mu$ increases.  The perturbative
approximation, however, fails long before reaching the
fixpoint. Therefore we examine two different two-loop cutoff
conditions: 
\begin{equation}
\lambda_c(\Lambda)=\lamFP/4 \qquad {\rm and} \qquad 
\lambda_c(\Lambda)=\lamFP/2\,.
\end{equation}  
The first
choice corresponds to a two-loop correction of 25\% to the one-loop
beta function $\beta_\lambda$, see Eq.~(\ref{betafct}). Perturbation
theory is expected to be reliable for such a value of
$\lambda(\Lambda)$ \cite{RW}.
The second choice causes a 50\% correction, and its value is
comparable with upper bounds on $\lambda(\Lambda)$ which can
be obtained from lattice calculations \cite{LW,aachen,heller}.
In addition, it is also 
relatively
close to the upper bound of the perturbative regime  
\cite{RW}.

\begin{figure}[tb]
\vspace*{30pt}
\centerline{
\epsfysize=2.56in \rotate[l]{\epsffile{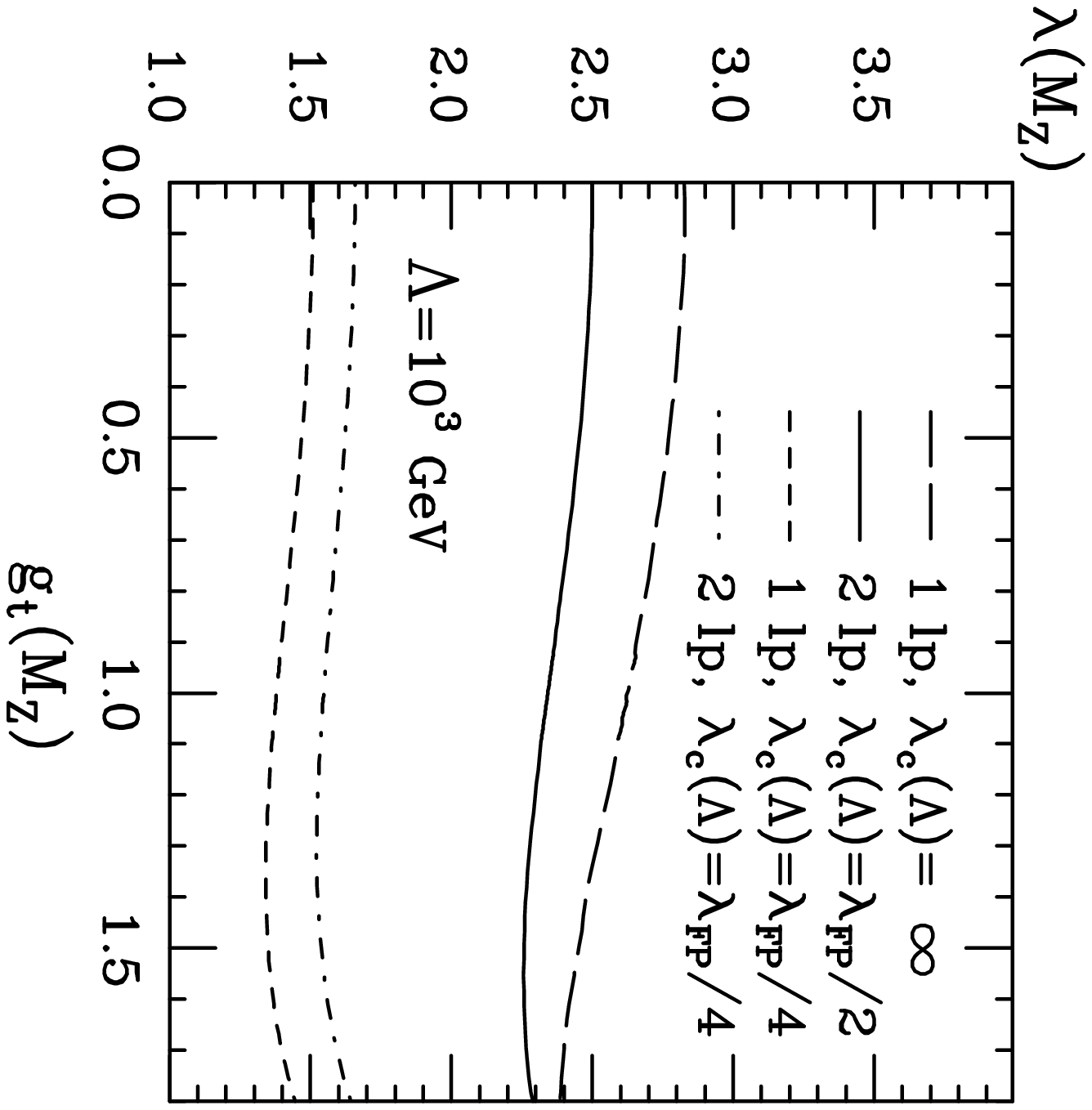}}
\hspace{1.0cm}
\epsfysize=3.02in \rotate[l]{\epsffile{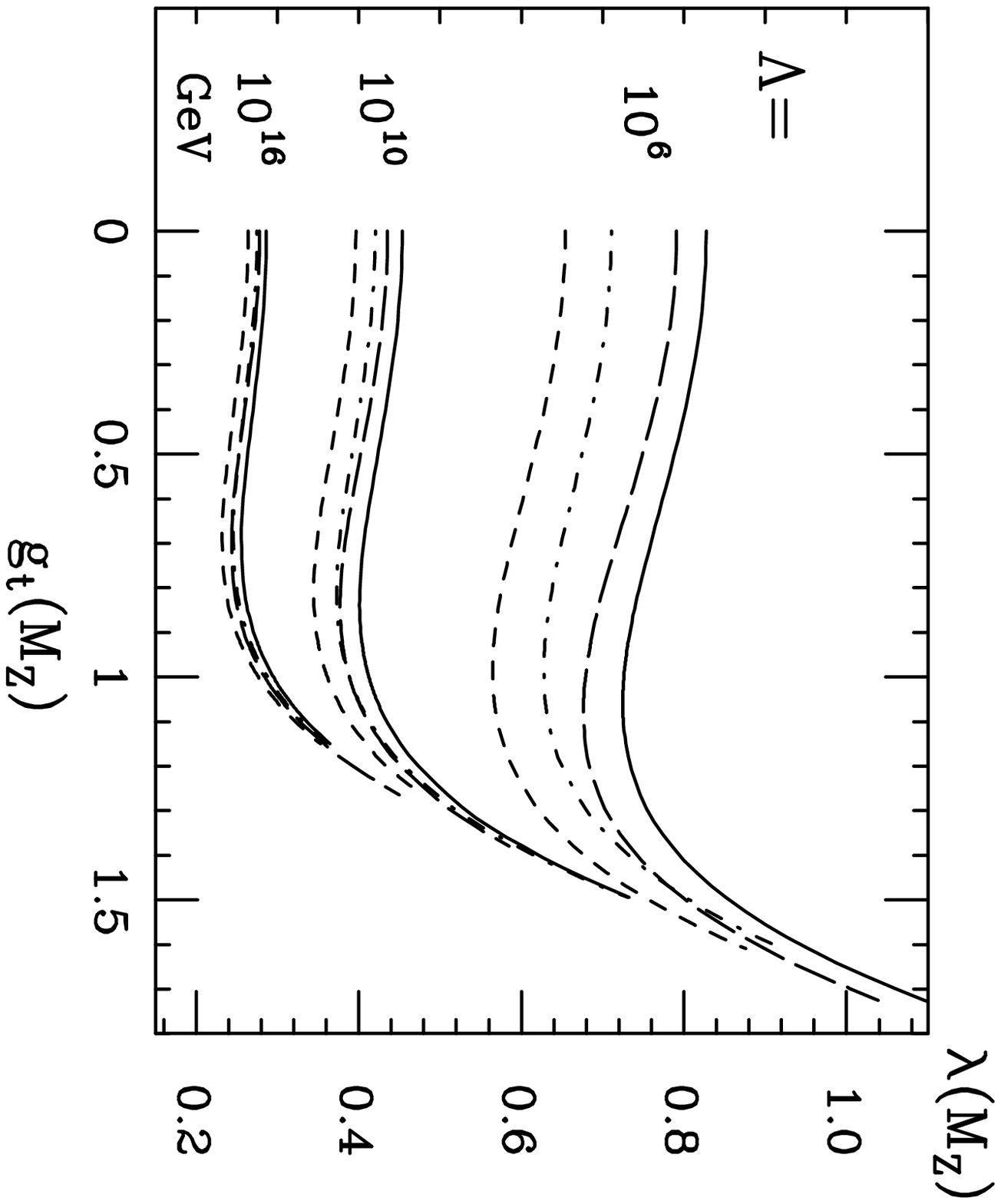}}
}
\vspace{0.5in}
\caption{Choosing either one-loop or two-loop RG evolution and various
  cutoff conditions $\lambda_c(\Lambda)$, the maximally allowed value of
  $\lambda(M_Z)$ is given as a function of $g_t(M_Z)$.  The cutoff condition
  $\lambda_c(\Lambda)$ is imposed at scales $\Lambda=10^3$ GeV (left plot) and
  $\Lambda=10^6,10^{10},10^{16}$ GeV (right plot). }
\label{figcutcond}
\end{figure}

Choosing four different embedding scales, $\Lambda=10^3$, $10^6$, $10^{10}$,
and $10^{16}$ GeV, we give in Fig.~\ref{figcutcond} the different values of
$\lambda(M_Z)$ which lead to the corresponding cutoff conditions
$\lambda_c(\Lambda)$ when evolving all SM couplings from $M_Z$ to $\Lambda$.
The one-loop result in Fig.~\ref{figcutcond} with $\lambda_c(\Lambda)=\infty$
agrees with the result obtained by Lindner \cite{lindner} when setting
$\lambda(M_Z)=M_H^2/2v^2$ and $g_t(M_Z)=\sqrt{2}m_t/v$, and taking into account
the updated experimental input for the various couplings at
$\mu_0=M_Z$.\footnote{ 
  In the case of $\Lambda=10^3$ GeV, for which Lindner \cite{lindner} only
  gives a qualitative estimate, we find a slightly higher upper bound on
  $\lambda(M_Z)$.}
The recent results in \cite{koreans} which question the Lindner results at all
scales $\Lambda$ are incorrect.\footnote{
  For large scale $\Lambda$, the errors in \cite{koreans} seem to be partially
  connected to the errorneous use of $10^n$ instead of ${\rm e}^n$ in all
  equations and figures where $\Lambda$ is specified.  This replacement,
  however, still does not correct all their results.}

Taking instead the value $\lambda_c(\Lambda)=\lamFP/4$ at one loop, we find a
value of $\lambda(M_Z)$ for which perturbation theory is definitely reliable
when evolving all SM couplings to $\Lambda$. For $\Lambda=10^{16}$ GeV the
one-loop perturbative upper bound on $\lambda(M_Z)$ is only slightly less than
the nonperturbative value obtained using the Landau pole criterion, indicating
the insensitivity of the upper bound to the cutoff condition. For
$\Lambda=10^3$ GeV, however, the perturbative upper bound is about 50\% less
than the Landau-pole bound, a sign for a strong dependence on the cutoff
condition $\lambda_c(\Lambda)$.

Going to two loops, the perturbative bound corresponding to
$\lambda_c(\Lambda)=\lamFP/4$ differs from the corresponding one-loop result by
less than 12\%: perturbation theory indeed seems applicable.  The maximal
upper bound as modelled by $\lambda_c(\Lambda)=\lamFP/2$ gives upper bounds on
$\lambda(M_Z)$ which are of the order of the one-loop Landau pole bounds.  

We conclude that our two-loop cutoff conditions are suitable for representing
two scenarios: $\lambda_c(\Lambda)=\lamFP/4$ corresponds to a perturbatively
reliable Higgs sector at embedding scale $\Lambda$, and the 
condition $\lambda_c(\Lambda)=\lamFP/2$ is at the verge of being
nonperturbative. 

The procedure for obtaining an $M_H$ upper bound from the bound
on $\lambda(M_Z)$ is as follows.
The couplings $\lambda(M_Z)$ and $g_t(M_Z)$ in
Fig.~\ref{figcutcond} are $\MSbar$ couplings at $\mu=M_Z$.  
The $\MSbar$ gauge couplings are
fixed at $M_Z$ using Eqs.~(\ref{alphamsbar}) and (\ref{sinmsbar}), and
$\alpha_s(M_Z)=0.118$.  The matching scale for the top quark coupling is taken
according to our previous discussion (Sect.~I) as $\mu_{0,t}\equiv m_t$, and we
take $m_t=175$ GeV. (Taking $\mu_{0,t}=M_Z$ has little effect on the final
numerical results.)  The matching scale for the Higgs coupling is chosen to be
$\mu_{0,H}\equiv M_H$ as argued above.  With these settings, we evolve
$\bar\lambda(M_Z)$ and all other SM couplings from $M_Z$ to some value
$\mu_{0,H}$ such that Eq.~(\ref{mchiggs}) is solved for some value $M_H$ with
$\mu_{0,H}=M_H$.  Subsequent evolution to $\mu_{0,t}$ checks the top quark
matching condition, Eq.~(\ref{mctop}), using $m_t=175$ GeV and the value of
$M_H$ found in the previous step.  If the top quark matching condition is not
satisfied, we iterate our procedure, starting at scale $M_Z$ with a different
value of $\bar g_t(M_Z)$.  Eventually, we find a final solution for $M_H$ which
is consistent with both matching conditions.
To investigate the importance of the one-loop matching corrections,
we repeat the above procedure taking the
matching corrections $\delta_H$ and $\delta_t$ to be zero.

\begin{figure}[tb]
\vspace*{30pt}
\centerline{
\epsfysize=3.0in \rotate[l]{\epsffile{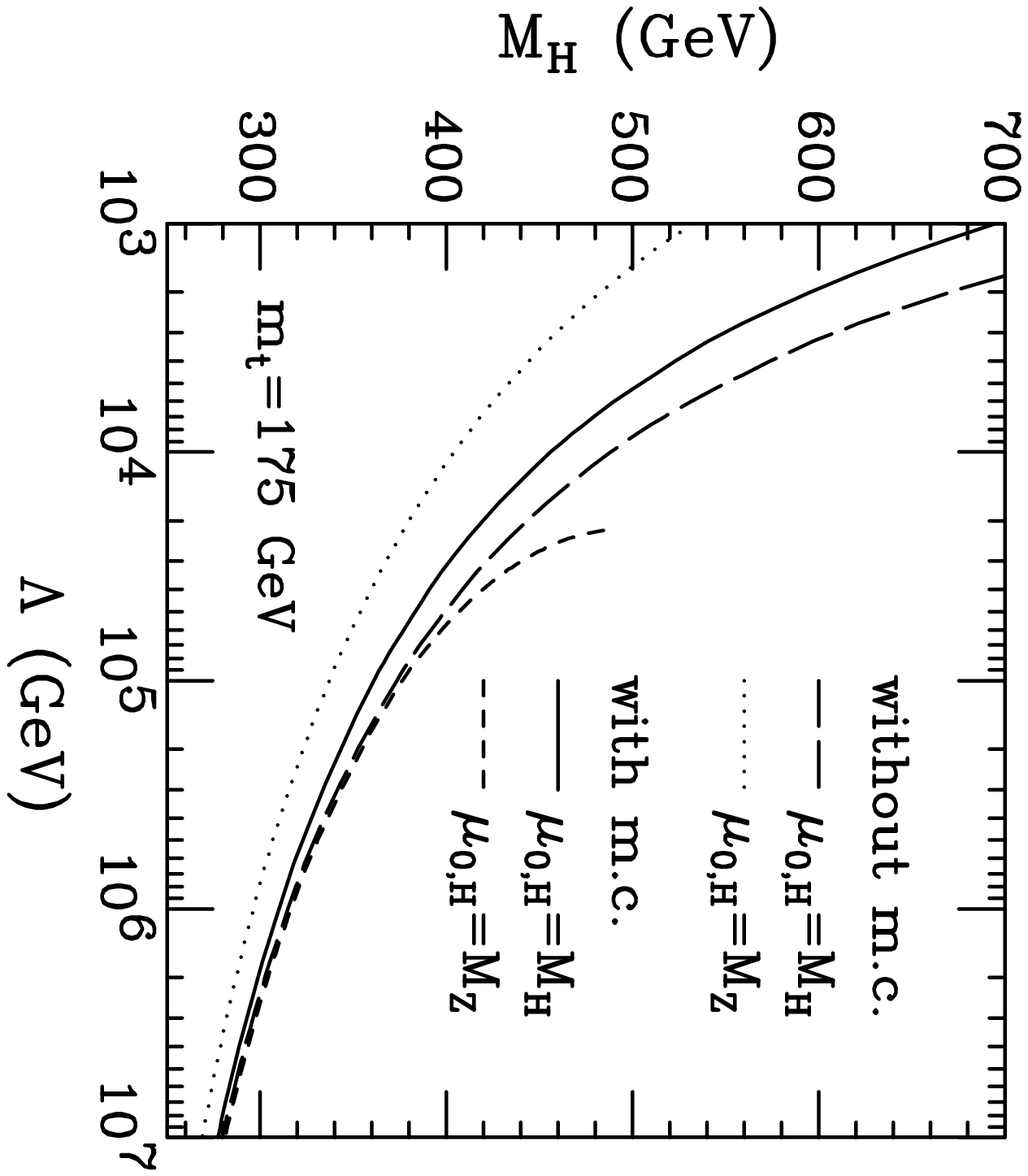}}
\hspace{0.8cm}
\epsfysize=2.95in \rotate[l]{\epsffile{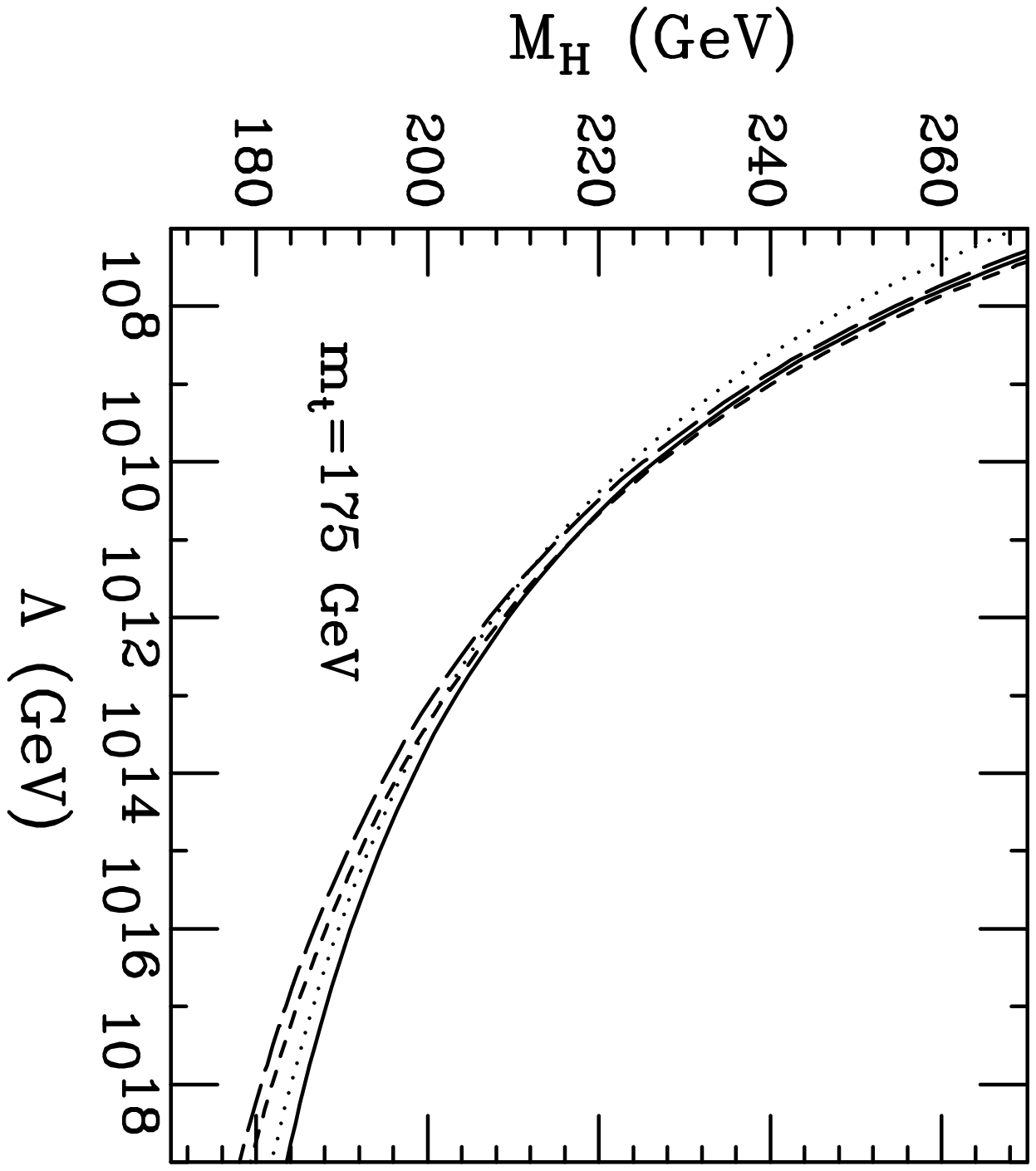}}
}
\vspace{0.5in}
\caption{Choosing two-loop RG evolution and cutoff condition
  $\lambda_c(\Lambda)=\lamFP/2$, the upper bound on $M_H$ is
  calculated. The running Higgs and Yukawa couplings, $\lambda(\mu)$
  and $g_t(\mu)$, are fixed by
  the physical masses $M_H$ and $m_t$ using matching conditions with
  and without one-loop matching corrections. In addition, the Higgs
  matching scale is varied to be $\mu_{0,H}=M_H$ and $M_Z$.  
  The top-quark mass is fixed
  at $m_t=175$ GeV, and $\mu_{0,t}=m_t$.  
  The left plot shows the result for small values
  of $\Lambda$, the right plot extends up to values of
  $\Lambda=10^{19}$ GeV.}
\label{figmudep}
\end{figure}

In Fig.~\ref{figmudep} we show the resulting two-loop upper bound on $M_H$ with
and without 
the use of matching corrections,
fixing the cutoff condition as
$\lambda_c(\Lambda)=\lamFP/2$.  Using the choice $\mu_{0,H}=M_H$,
the comparison of the solid line (with matching corrections) 
and long-dashed line (without matching corrections)
allows for a conservative estimate of higher order corrections. We find
that the difference of the two results can exceed 100 GeV at small embedding
scale $\Lambda$, but reduces to less than about 6 GeV at large scale. 

In addition to the preferred choice $\mu_{0,H}=M_H$,
we also give results when using $\mu_{0,H}=M_Z$.  
For large
embedding scale $\Lambda$ (resulting in small values of $M_H$), the two
different choices of $\mu_{0,H}$ give similar results.  For small scale
$\Lambda$, the difference is significant (Fig.~\ref{figmudep}, dotted line).
This was already anticipated in a one-loop study of pure $\phi^4$ theory
which underlies the SM Higgs sector \cite{HHG}.
However, the inclusion of matching corrections (short-dashed curve) shows that
the scale choice $\mu_{0,H}=M_Z$ is completely inadequate for large values
of $M_H$ as indicated by the largeness of the corrections compared to
the choice $\mu_{0,H}=M_H$.
Even more strikingly,
values $\Lambda<2\times10^4$ GeV (which lead to bounds $\lambda(M_Z)>1.2$
when using $\lambda_c(\Lambda)=\lamFP/2$)
have no solution in $M_H$ which satisfy the $\MSbar$ matching condition.
This is due to the fact that the choice $\mu_{0,H}=M_Z$ restricts the $\MSbar$
coupling to a maximal value $\bar\lambda(M_Z)=1.2$ which is obtained for
$M_H\approx 495$ GeV.  We will only consider the $\mu_{0,H}=M_H$
results when determining the final $M_H$ upper bounds.

The results of Fig.~\ref{figmudep} can also be compared with the two-loop
results of \cite{grad}.  There no matching corrections have been included, and
$\mu_{0,H}=M_Z$ is used.  The cutoff-condition $\lambda_c(\Lambda)$ is
determined as a turning point in the two-loop calculation rather than a fixed
value.  This procedure yields larger two-loop values of $\lambda_c(\Lambda)$
than used here.  The resulting $M_H$ bounds are therefore larger than our
corresponding result with $\mu_{0,H}=M_Z$ and no matching corrections, but with
$\lambda_c(\Lambda)=\lamFP/2$.

\begin{figure}[tb]
\vspace*{30pt}
\centerline{
\epsfysize=3.8in \rotate[l]{\epsffile{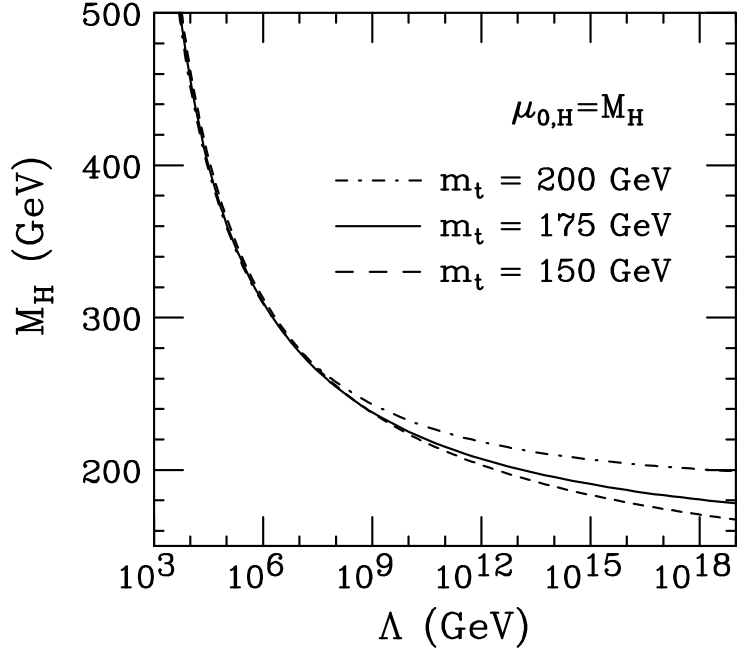}}
}
\vspace{0.5in}
\caption{The dependence of the upper $M_H$ bound on the top-quark
  mass.  The $\MSbar$ matching conditions with $\mu_{0,H}=M_H$ and
  $\mu_{0,t}=m_t$ 
  are used in connection with two-loop RG evolution and cutoff
  condition $\lambda_c(\Lambda)=\lamFP/2$.  For low values of the
  embedding scale $\Lambda$, the $M_H$ upper bound is insensitive to
  the exact value of $m_t$.  For large embedding scales there is a
  larger $m_t$ dependence. 
  Without matching corrections (not shown), the top mass dependence is 
  qualitatively the same.}
\label{figmtdep}
\end{figure}

In Fig.~\ref{figmtdep} we analyse the dependence of the upper $M_H$
bound on $m_t$. Varying $m_t$ in the range 150--200 GeV, the bound on
$M_H$ changes less than 40 GeV for the largest embedding scale
considered, $\Lambda= 10^{19}$ GeV. 
The latest experimental
result \cite{warsawMT}, $m_t=175\pm6$ GeV, reduces this
uncertainty to less than 5 GeV at the $1\sigma$ level.  For embedding scales
$\Lambda<10^{10}$ GeV the uncertainty due to $m_t$ can then entirely be
neglected compared to the theoretical uncertainties connected to the
cutoff condition and higher-order corrections. The uncertainty
in the QCD coupling, $\alpha_s(M_Z)=0.118\pm0.003$ \cite{alphaqcd}, causes
a shift of less than 1 GeV in the $M_H$ upper bound, with the maximal effect
at $\Lambda=10^{19}$ GeV. 

\begin{figure}[tb]
\vspace*{30pt}
\centerline{
\epsfysize=3.8in \rotate[l]{\epsffile{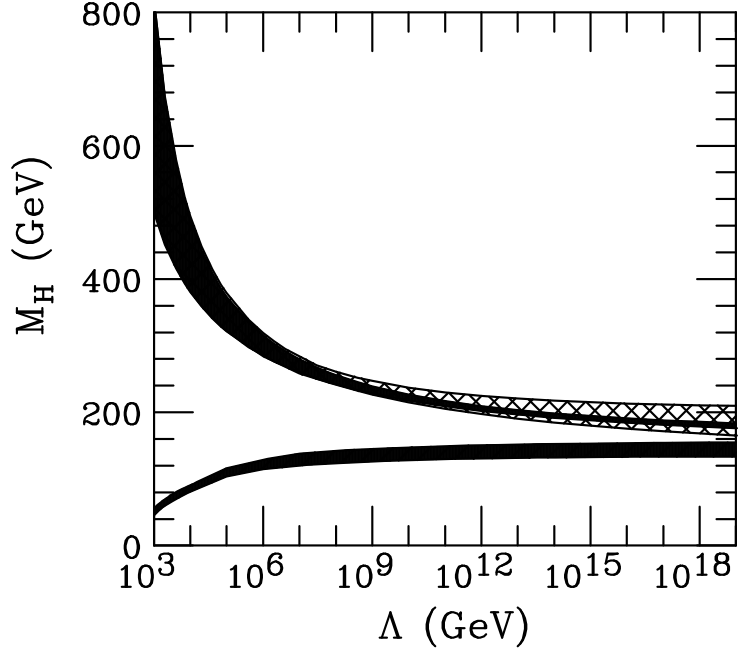}}
}
\vspace{0.5in}
\caption{Summary of the uncertainties connected to the bounds on $M_H$.
  The upper solid area indicates the sum of theoretical
  uncertainties in the $M_H$ upper bound when keeping $m_t=175$ GeV
  fixed. The cross-hatched 
  area shows the additional uncertainty when varying $m_t$ from 150 to
  200 GeV. 
  The upper edge corresponds to Higgs
  masses for which the SM Higgs sector ceases to be
  meaningful at scale $\Lambda$ (see text), and the 
  lower edge indicates a value of
  $M_H$ for which perturbation theory is certainly expected to be
  reliable at scale $\Lambda$.
  The lower solid area represents the theoretical uncertaintites in
  the $M_H$ lower bounds derived from stability requirements
  \protect\cite{altisi,casas1,casas2} using $m_t=175$ GeV and
  $\alpha_s=0.118$.} 
\label{figfinal}
\end{figure}

In summary, we have discussed the uncertainties in the $M_H$ upper
bound due to the choice of the cutoff condition
(Fig.~\ref{figcutcond}),  
the importance of one-loop matching corrections
and the choice of the matching scale $\mu_{0,H}$
(Fig.~\ref{figmudep}), and the top-quark mass
dependence (Fig.~\ref{figmtdep}). 
Fixing the top quark mass to be 175
GeV, using two-loop beta functions and appropriately 
choosing the matching scale to be
$\mu_{0,H}=M_H$, we find the sum of all
theoretical uncertainties to be represented by the upper solid area
indicated in Fig.~\ref{figfinal}. They are obtained by choosing
$\mu_{0,H}=M_H$ and using
matching conditions with and without one-loop
matching corrections. The cutoff condition is varied 
between $\lambda_c(\Lambda)=\lamFP/4$ and
$\lamFP/2$.  
The lower edge of the solid area indicates a
value of $M_H$ for which perturbation theory is certainly reliable up to 
scale $\Lambda$; in particular, 
the triviality problem of the standard model is clearly
avoided for such values of
$\Lambda$ and $M_H$. The upper edge of the solid area can be used
to estimate the scale $\Lambda(M_H)$ at which the standard model 
ceases to be meaningful as an effective theory. Although the perturbative
approach does not allow for extraction of absolute upper bounds, 
the consideration of lattice calculations in $\phi^4$ theory 
seems to reinforce or even tighten the upper bounds presented here
\cite{LW,heller,aachen,RW}. 
For low values of $\Lambda$, 
the one-loop Landau pole bounds of \cite{lindner}
are found to be near the perturbative lower edge
of the upper solid area in Fig.~\ref{figfinal}.
The additional experimental uncertainty due to the top
quark mass is represented 
by the cross-hatched area in Fig.~\ref{figfinal}, generously varying the
top-quark mass from 150 GeV to 200 GeV. The present-day $1\sigma$ result of 
$m_t=175\pm6$ GeV is sufficient to make it the smallest source of
error except for large values of the embedding scale $\Lambda$. 
In particular, we find:
\begin{equation}
M_H < 180 \pm 4\pm 5 \;{\rm GeV} \qquad 
{\rm if} \qquad \Lambda=10^{19}\; {\rm GeV}, 
\end{equation}
the first error indicating the theoretical
uncertainty, the second error reflecting the $m_t$
dependence.\footnote{The very recent result \cite{isidori} of $M_H<174$ GeV
  for $\Lambda=10^{19}$ GeV is lower than our lowest result due to the use of
  the smaller cutoff condition $\lambda_c(\Lambda)=5/3<\lamFP/4\approx3$ (our
  notation).} 

For comparison, we also give the lower bounds on $M_H$ from stability
conditions on the SM Higgs effective potential. At large scale $\Lambda$, the
stability bound is well approximated by requiring the Higgs running coupling to
remain positive: $\lambda(\Lambda)>0$. Such an analysis has been carried out at
the two-loop level including matching corrections \cite{altisi}, and they agree
within the theoretical errors with a more careful treatment of the one-loop
effective potential \cite{casas1}. The discrepancy at scales $\Lambda<10$ TeV
has been resolved recently \cite{casas2}, and we use the latter
results.  Fixing $m_t=175$ GeV and
$\alpha_s(M_Z)=0.118$ we show the lower bound in Fig.~\ref{figfinal}
(lower solid area), with
the solid area indicating the theoretical uncertainty.  At large $\Lambda$, the
theoretical error is estimated by using $\mu_{0,H}=m_t$ and comparing the
results with and without matching corrections, and at low $\Lambda$ the
theoretical error is $\pm 5$ GeV according to \cite{casas2}.  The variation
$m_t=175\pm25$ GeV yields a much larger uncertainty in the $M_H$ lower bound
than in the $M_H$ upper bound and is not shown.

\vspace{15mm}

\noindent{\bf Acknowledgements}\\

\noindent The authors thank L.~Durand, M.~Lindner and J.~Pestieau for valuable
discussions and reading of the manuscript.  One author (T.H.)  thanks the
Bundes\-ministerium f\"ur Forschung und Technologie (BMFT) for support under
contract no.\ 056DO93P(5).  The other author (K.R.) acknowledges support by
the Deutsche Forschungs\-gemein\-schaft (DFG) under contract no.\
DFG-Li-519/2-1, and is grateful for the hospitality of the University of
Wisconsin -- Madison where parts of this work were completed.

\newpage
\begin{center} 
{\Large{\bf Figures}}
\end{center}

\vspace{1cm}

\newpage

\newpage

\vspace{1cm}

\newpage

\vspace{1cm}

\newpage

\vspace{1cm}

\newpage

\vspace{1cm}

\end{document}